# ENHANCING CPU PERFORMANCE USING SUBCONTRARY MEAN DYNAMIC ROUND ROBIN (SMDRR) SCHEDULING ALGORITHM

Sourav Kumar Bhoi[*1], Sanjaya Kumar Panda[2] and Debashee Tarai[3]
[*1]Department of Computer Science & Engineering, National Institute of Technology, Rourkela, Odisha, India
souravbhoi@gmail.com[1]
[2]Department of Computer Science & Engineering, National Institute of Technology, Rourkela, Odisha, India
sanjayauce@gmail.com[2]
[3]Department of Computer Science & Engineering, Indian Institute of Technology, Bombay, Maharashtra, India
debashee.tarai@gmail.com[3]

***Abstract:*** Round Robin (RR) Algorithm is considered as optimal in time shared environment because the static time is equally shared among the processes. If the time quantum taken is static then it undergoes degradation of the CPU performance and leads to so many context switches. In this paper, we have proposed a new effective dynamic RR algorithm SMDRR (Subcontrary Mean Dynamic Round Robin) based on dynamic time quantum where we use the subcontrary mean or harmonic mean to find the time quantum. The idea of this approach is to make the time quantum repeatedly adjusted according to the burst time of the currently running processes. Our experimental analysis shows that SMDRR performs better than RR algorithm in terms of reducing the number of context switches, average turnaround time and average waiting time.

## INTRODUCTION

Operating systems are primarily resource managers; the main resource they manage is computer hardware in the form of processors, storage, input/output devices, communication devices and data [1]. Operating system performs many functions such as implementing the user interface, sharing hardware among users, allowing users to share data among themselves, preventing users from interfering with one another, scheduling resources among users, facilitating input/output, recovering from errors, accounting for resource usage, facilitating parallel operations, organizing data for secure and rapid access, and handling network communications [1]. As we know RR algorithm is the mostly used algorithm in time sharing system for its simplicity. RR algorithm gives equal priority to each process by sharing a common time quantum (TQ).

We know that CPU scheduling is the task of selecting a waiting process from the ready queue and allocating the CPU to it [2]. The performance of the CPU mainly depends upon many criteria's such as *CPU utilization, Throughput, Turnaround Time (TAT), Waiting Time (WT), Context Switch (CS), Response Time* etc. The utilization of the CPU is called *CPU utilization* where we keep the CPU as busy as possible. The number of processes completed per unit time is called *Throughput. Waiting Time* is the sum of the periods spent waiting in the ready queue [2]. Time from the submission of a request until the first response is produced is called *Response Time*. *Turnaround Time* is the interval from the time of the submission of a process to the time of completion is the turnaround time. *Context switch* is the number of times the process switches to get execute. In this paper, we have proposed a new effective dynamic RR algorithm SMDRR (Subcontrary Mean Dynamic Round Robin) based on dynamic time quantum where we use the subcontrary mean or harmonic mean to find the time quantum for the processes to execute. The idea of this approach is to make the time quantum repeatedly adjusted according to the burst time of the currently running processes. Our experimental analysis shows that SMDRR performs better than RR algorithm in terms of reducing the number of context switches, average waiting time and average turnaround time.

### *Process Scheduling Algorithms:*

CPU scheduling deals with the problem of deciding which of the processes in the ready queue is to be allocated to the CPU [2]. Moreover we should distinguish between the two schemes of scheduling: preemptive and non-preemptive algorithms. Preemptive algorithms are those where the burst time of the process being in execution is preempted when a higher priority process arrived [3]. Non-preemptive algorithms are used where the process runs to complete its burst time even a higher priority process arrives during its execution time [3]. There are many scheduling policies for executing the processes. *First come first served* is the simplest scheduling algorithm which queues processes in the order that they arrive in the ready queue. *Shortest job first* arranges processes with the least estimated processing time remaining to be next in the queue. *Fixed priority pre emptive scheduling* is commonly used in real-time systems.

With fixed priority pre-emptive scheduling, the scheduler ensures that at any given time, the processor executes the highest priority task of all those tasks that are currently ready to execute. *Round-robin scheduling* Round-robin (RR) is one of the simplest scheduling system assigns time slices to each process in equal portions and in circular order, handling all the processes without priority (also known as cyclic executive). Round-robin scheduling is both simple and easy to implement and starvation free. *Earliest Deadline First* (EDF) places processes in a priority queue. Whenever a scheduling event occurs (task finishes, new task released, etc.) the queue will be searched for the process closest to its deadline. This process is the next to be scheduled for execution. These are some of the process scheduling algorithms for execution of the processes.

***Previous Work Done:***

Many research works has been done under this topic to enhance the performance of CPU. The static time quantum which is a limitation of RR was removed by taking dynamic time quantum by Matarneh [4]. Priority Based Dynamic Round Robin Algorithm (PBDRR), which calculates intelligent time slice for individual processes and changes after every round of execution [5]. A new RR scheduling algorithm is developed by taking the mean as the time quantum [3].

## SMDRR ALGORITHM

In our proposed algorithm, the time quantum is taken as the subcontrary or harmonic mean of the increasingly sorted burst time of all the processes and this change dynamically in every cycle till the end of processes.

***Method Taken:***

In our algorithm, the processes are first sorted in ascending order of their burst time to give better context switch, turnaround time and waiting time. Performance of RR algorithm solely depends upon the size of time quantum taken. If it is very small, it causes too many context switches and lowers the CPU efficiency. If it is very large, gives poor response to short interactive requests. So our algorithm solves this problem by taking a dynamic time quantum where the time quantum is repeatedly adjusted according to the remaining burst time of currently running processes. To get the optimal time quantum, subcontrary mean or harmonic means of the burst time of the processes are taken as the time quantum.

***Proposed Algorithm:***

In our algorithm, when processes are already present in the ready queue, their arrival time is assigned to zero before they are allocated to the CPU. The burst time and the number of processes (n) are accepted as input with a counter value 'i'. Let TQ be the time quantum. We calculate the time quantum by using the formula as follows (1):

Subcontrary Mean (SM) = $n / ( 1/x_1 + 1/x_2 + .. + 1/x_n)$ (1)

Where n = Total no. of processes

X = Set of processes

and $(x_1, x_2, ..... , x_n) \in X$

In this algorithm we first sort the processes according to their burst time. When the ready queue is not empty, we calculate the TQ as the subcontrary mean or harmonic mean to execute the processes. After this we calculate the remaining burst time of the processes and then check different conditions in step 5 of our proposed algorithm show in figure 1. Then after the completion of step 5 we go for step 6 where we calculate the Average Turnaround Time (ATT), Average Waiting Time (AWT), number of Context Switches (CS). Then go to step 7 for end of the process. The flowchart for the algorithm is shown in figure 2.

1. First all the processes present in ready queue are sorted in ascending order of their burst time.
   n → number of processes
   i → counter value

2. While(RQ!= NULL)
   //TQ = Subcontrary Mean (remaining burst time of all the processes)

   TQ = $n / ( 1/x_1 + 1/x_2 + .. + 1/x_n )$

   // n = Total no. of processes
   // X = Set of processes ,where $(x_1,x_2.....x_n) \in X$
   //RQ = Ready Queue
   //TQ = Time Quantum

3. Assign TQ to (1 to n) process
   for i = 1 to n
   {
   $P_i$ → TQ
   }
   end for
   // Assign TQ to all the available processes.

4. Calculate the remaining burst time of the processes.

5. if ( new process is arrived and BT != 0 )
   go to step 1
   else if ( new process is not arrived and BT != 0 )
   go to step 2
   else if ( new process is arrived and BT == 0)
   go to step 1
   else
   go to step 6
   end if
   end while

6. Calculate ATT, AWT and CS.
   //ATT = Average Turnaround Time
   //AWT = Average Waiting Time
   //CS = number of Context Switches

7. End

Figure. 1: Pseudocode of SMDRR algorithm

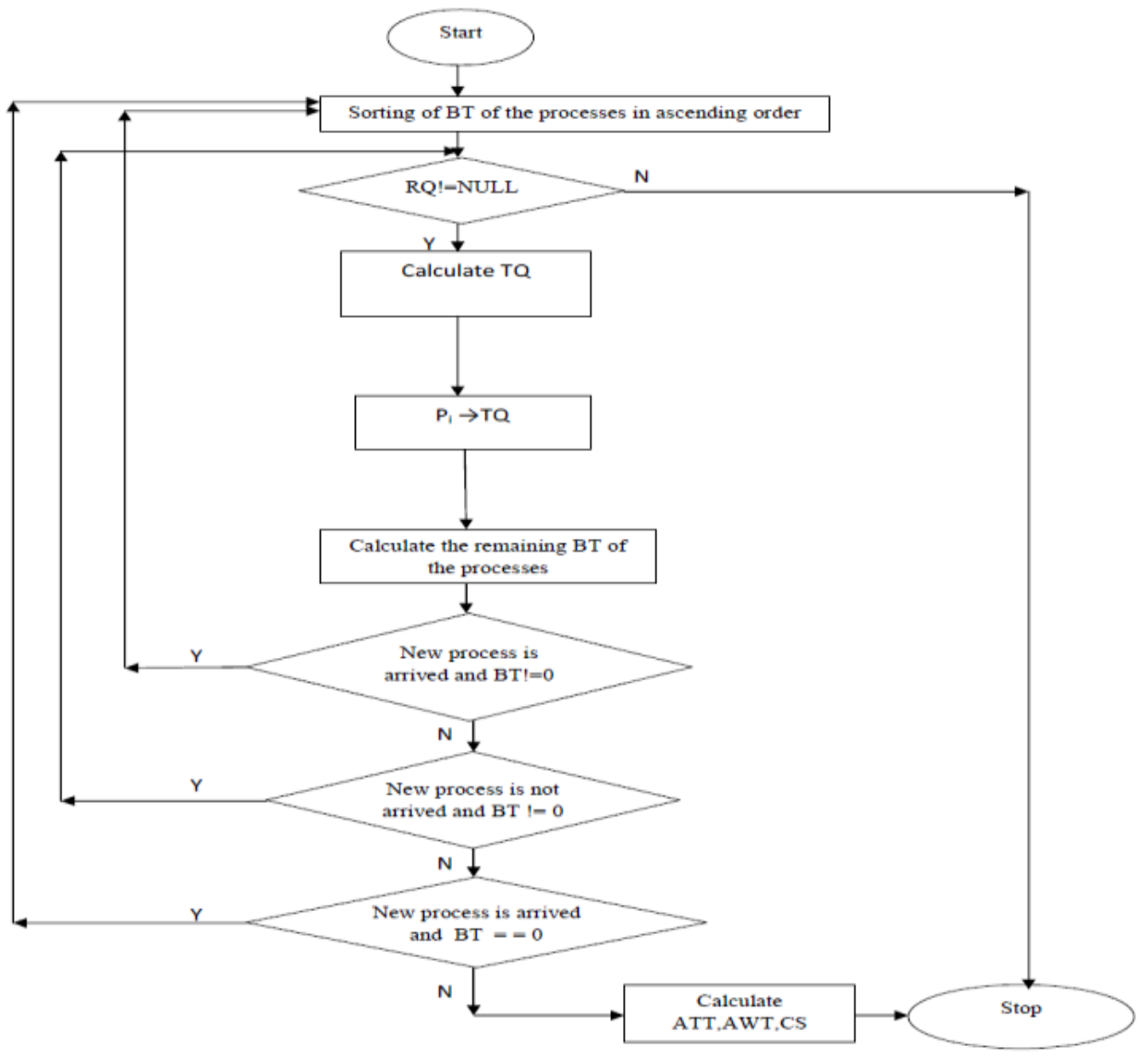

Figure. 2: Flowchart for SMDRR Algorithm

***Illustration:***

Given the burst time sequence of the processes as P1= 20, P2= 40, P3= 83, P4= 90. Initially the burst time of all the processes were sorted in ascending order which resulted in sequence P1, P2, P3, P4. Then the harmonic mean of the above burst time which was calculated to be 41 (ceiling value) was assigned as the time quantum for all the processes. In the next step remaining burst time of each process was calculated after allocating the time quantum. After first cycle the remaining burst time sequence for above processes changed to P2=42 and P3=49. When a process completes its burst time, it gets deleted from the ready queue automatically. So in this case, the processes P1 and P2 were deleted from the ready queue. The present remaining burst time were sorted in increasing order and then the harmonic mean of these burst times was assigned as the time quantum where we get 46 as the time quantum for the second cycle. The remaining burst time after the second cycle is P4=3, then we calculate the harmonic mean as 3 and assign it to P4.This is how the processes are executed in the ready queue. The above process was continued till all the processes were deleted from the ready queue.

## *PERFORMANCE EVALUATION*

### *Assumptions Taken:*

All the processes are independent of each other. There is equal priority to all the processes. All the processes are done on a single processor environment. All the attributes like burst time, number of processes and the time slice of all the processes are known before submitting the processes to the processor. All processes are CPU bound. No processes are I/O bound. Since, the cases are assumed to be close to ideal, the Context Switching Time is equal to zero i.e. there is no Context Switch Overhead incurred in switching from one process to another. The TQ is taken in milliseconds (ms).

### *Performance Parameters:*

The significance of our performance parameters for experimental analysis is as follows:

- ***a.*** ***Turnaround time (TAT)*****:** For the better performance of the algorithm, average turnaround time should be less.
- ***b.*** ***Waiting time (WT)*****:** For the better performance of the algorithm, average waiting time should be less.
- ***c.*** ***Number of Context Switches (CS)*****:** For the better performance of the algorithm, the number of context switches should be less.

### *Experiments Performed*:

To evaluate the performance of our proposed algorithm, we have taken a pair of four processes and five processes in four different cases. Here for simplicity, we have taken four and five processes. The algorithm works effectively even if it used with a very large number of processes. In each case, we have compared the experimental results of our proposed algorithm with the round robin scheduling algorithm with fixed time quantum Q. Here we have assumed a constant time quantum TQ equal to 20 ms in all the four cases.

**Case 1:** We Assume four processes arriving at time = 0, with burst time (P1 = 20, P2 = 40, P3 = 83, P4 = 90). The Table-1 and Table-2 shows the processes with burst time and comparison of RR and SMDRR respectively. Figure-3 and Figure-4 shows the gantt chart of the two algorithms respectively.

Table 1. Processes with Burst Time (Case-1)

| *Process* | *Arrival Time* | *Burst Time* |
|---|---|---|
| P1 | 0 | 20 |
| P2 | 0 | 40 |
| P3 | 0 | 83 |
| P4 | 0 | 90 |

Table 2. Comparison of RR and SMDRR (Case-1)

| *Algorithm* | *TQ* | *TAT* | *WT* | *CS* |
|---|---|---|---|---|
| RR | 20 | 144 | 85.75 | 12 |
| SMDRR | 41,46,3 | 124.5 | 66 | 6 |

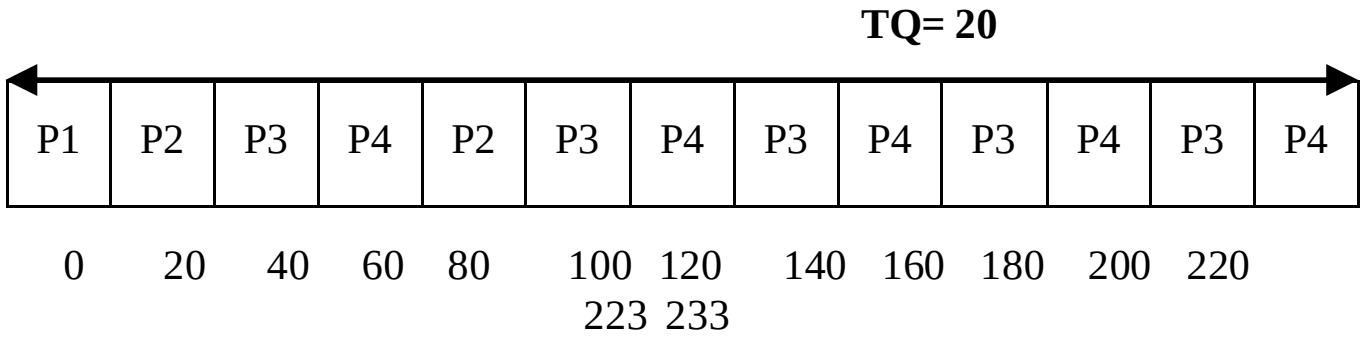

Figure: 3 Gantt Chart for RR (Case-1)

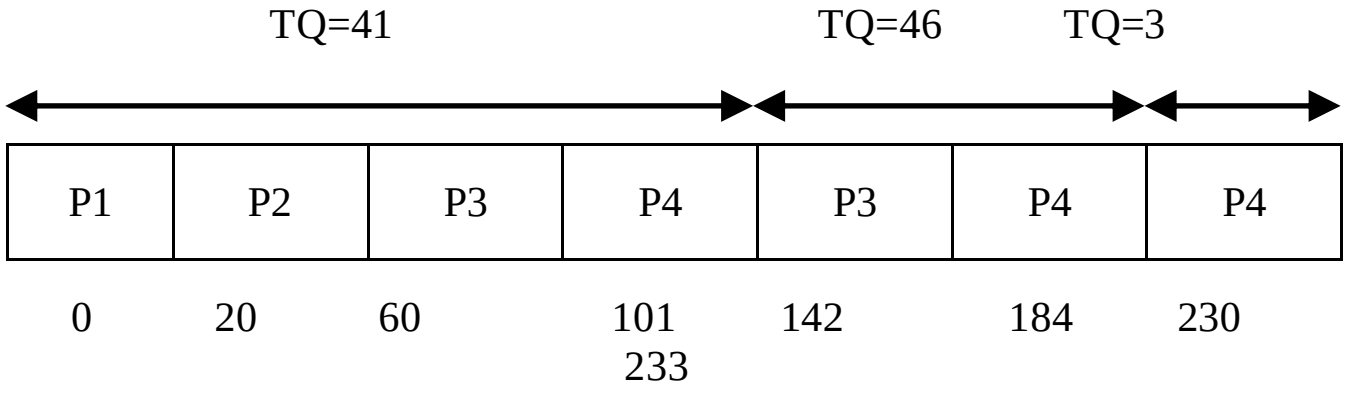

Figure. 4: Gantt chart for SMDRR (Case-1)

**Case 2:** We Assume five processes arriving at time = 0, with burst time (P1 = 17, P2 = 27, P3 = 52, P4 = 57, P5=59). The Table-3 and Table-4 shows the processes with burst time and the comparison of RR and SMDRR respectively. Figure-5 and Figure-6 shows the gantt chart of the two algorithms respectively.

Table 3. Processes with Burst Time (Case-2)

| *Process* | *Arrival Time* | *Burst Time* |
|---|---|---|
| P1 | 0 | 17 |
| P2 | 0 | 27 |
| P3 | 0 | 52 |
| P4 | 0 | 57 |
| P5 | 0 | 59 |

Table 4. Comparison of RR and SMDRR (Case-2)

| *Algorithm* | *TQ* | *TAT* | *WT* | *CS* |
|---|---|---|---|---|
| RR | 20 | 140.4 | 98 | 11 |
| SMDRR | 34,20,4,1 | 128.6 | 86.2 | 10 |

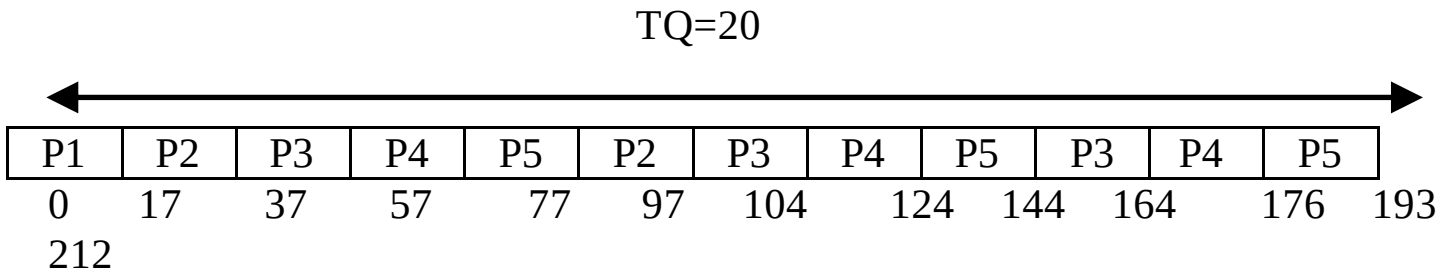

Figure. 5: Gantt chart for RR (Case-2)

TQ=34 TQ=20 TQ=4 TQ=1

| P1 | P2 | P3 | P4 | P5 | P3 | P4 | P5 | P4 | P5 | P5 |
|---|---|---|---|---|---|---|---|---|---|---|

0 17 44 78 112 146 164 184 204 207 211
212

Figure. 6: Gantt chart for SMDRR (Case-2)

**Case 3:** We Assume four processes with burst time P1 = 10, P2 = 14, P3 = 69, P4 = 75 and assumed the arrival time of P1=0, P2= 6, P3=12, P4=22. The Table-5 and Table-6 shows the processes with arrival time and burst time and comparison of RR and SMRR respectively. Figure-7 and Figure-8 shows the gantt chart of the two algorithms respectively.

Table 5. Processes with Burst Time and Arrival Time (case-3)

| *Process* | *Arrival time* | *Burst Time* |
|---|---|---|
| P1 | 0 | 10 |
| P2 | 6 | 14 |
| P3 | 12 | 69 |
| P4 | 22 | 75 |

Table 6. Comparison of RR and SMDRR (Case-3)

| *Algorithm* | *TQ* | *TAT* | *WT* | *CS* |
|---|---|---|---|---|
| RR | 20 | 88.75 | 47.75 | 9 |
| SMDRR | 10,14,72,3 | 73.75 | 32.75 | 4 |

TQ=20

| P1 | P2 | P3 | P4 | P3 | P4 | P3 | P4 | P3 | P4 |
|---|---|---|---|---|---|---|---|---|---|

0 10 24 44 64 84 104 124 144 153
168

Figure. 7: Gantt chart for RR (Case-3)

TQ=10 TQ=14 TQ=72 TQ=3

| P1 | P2 | P3 | P4 | P4 |
|---|---|---|---|---|

0 10 24 93 165
168

Figure. 8: Gantt chart for SMDRR (Case-3)

**Case 4:** We Assume five processes with burst time P1 = 18, P2 = 20, P3 = 50, P4 = 60, P5=68 and assumed the arrival time of P1=0, P2= 3, P3=6, P4=11, P5=21. The Table-7 and Table-8 shows the processes with arrival time and burst time and comparison of RR and SMRR respectively. Figure-9 and Figure-10 shows the gantt chart of the two algorithms respectively.

Table 7. Processes with Burst Time and Arrival Time (Case-4)

| *Process* | *Arrival time* | *Burst Time* |
|---|---|---|
| P1 | 0 | 18 |
| P2 | 3 | 20 |
| P3 | 6 | 50 |
| P4 | 11 | 60 |
| P5 | 21 | 68 |

Table 6. Comparison of RR and SMDRR (Case-4)

| *Algorithm* | *TQ* | *TAT* | *WT* | *CS* |
|---|---|---|---|---|
| RR | 20 | 125.6 | 82.4 | 11 |
| SMDRR | 18,35,25,43 | 108.6 | 65.4 | 7 |

TQ=20

| P1 | P2 | P3 | P4 | P5 | P3 | P4 | P5 | P3 | P4 | P5 | P5 |
|---|---|---|---|---|---|---|---|---|---|---|---|

0 18 38 58 78 98 118 138 158 168 188 208
216

Figure. 9: Gantt chart for RR (Case-4)

TQ=18 TQ=35 TQ=25 TQ=43

| P1 | P2 | P3 | P4 | P3 | P4 | P5 | P5 |
|---|---|---|---|---|---|---|---|

0 18 38 73 108 123 148 173
216

Figure. 10 : Gantt chart for SMDRR (Case-4)

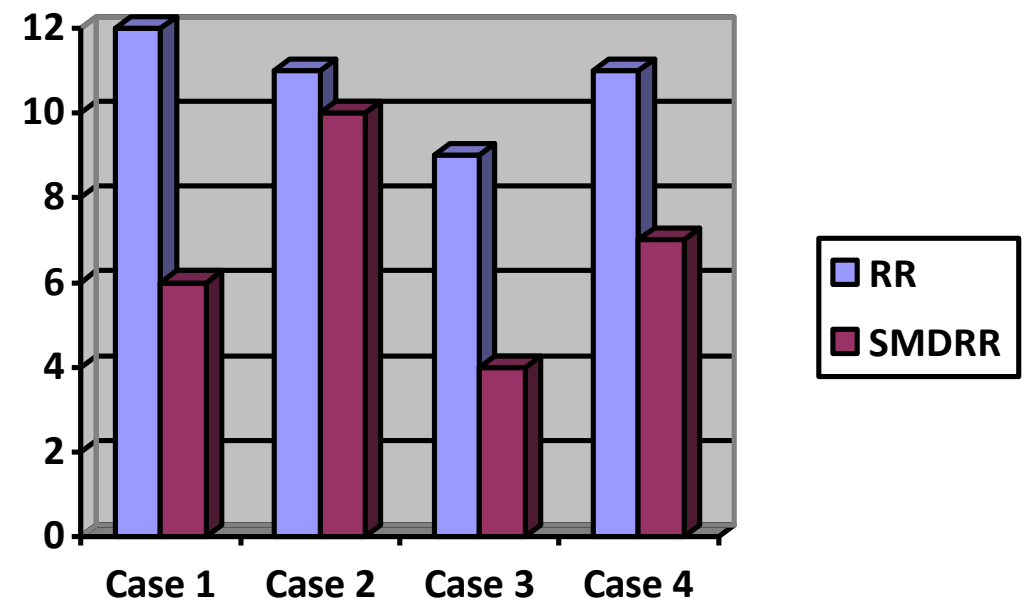

Figure.11: Comparison of Context Switch by taking static and dynamic time quantum for case1, case 2, case3 and case 4.

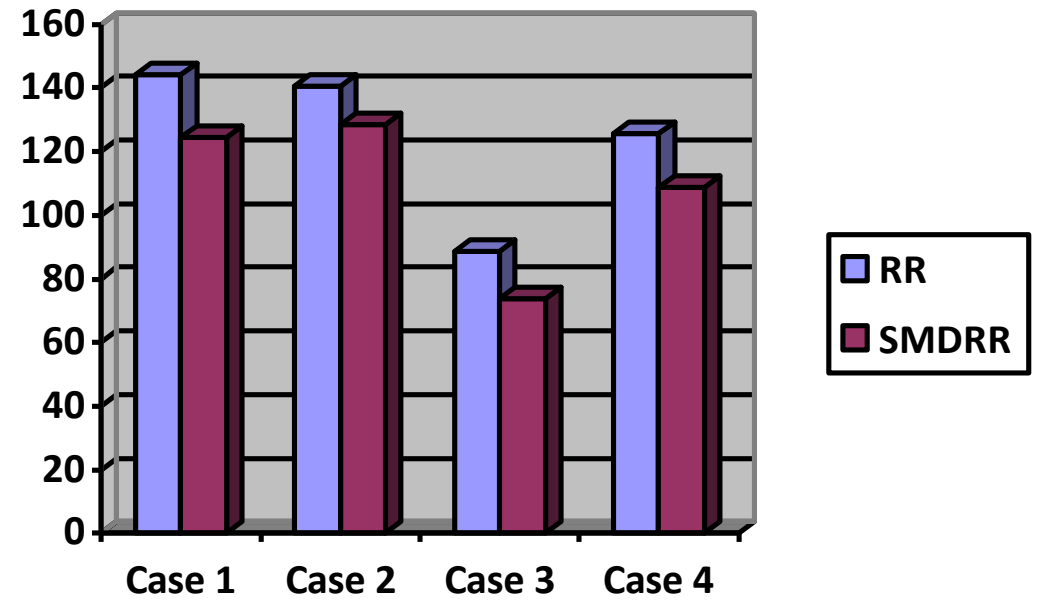

Figure.12: Comparison of Turnaround Time by taking static and dynamic time quantum for case1, case 2, case3 and case 4.

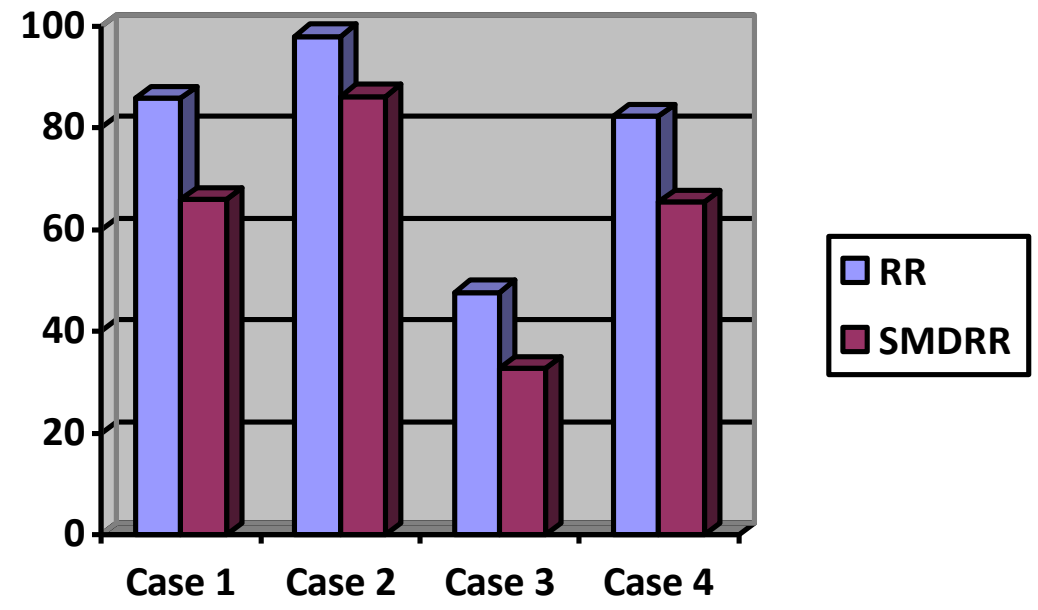

Figure. 13: Comparison of Waiting Time by taking static and dynamic tine quantum for case 1, case 2, case 3, case 4.

## CONCLUSION

From the above experiments, SMDRR algorithm shows better results than RR algorithm in enhancing the CPU

performance and its efficiency. By using our algorithm we are getting better Average Turnaround Time, Average Waiting Time and Context Switch. As we have taken the ideal cases in calculating the TAT, WT and CS .In future we can implement this algorithm in real time operating systems.

## Short Bio Data for the Author

Sourav Kumar Bhoi has received his B. Tech degree in Computer Science & Engineering from Veer Surendra Sai University of Technology, Odisha, India in 2011 and currently pursuing M. Tech degree in Computer Science & Engineering at National Institute of Technology, Rourkela, Odisha, India.

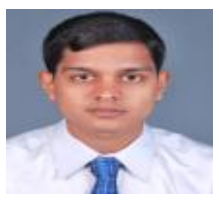

Sanjaya Kumar Panda has received his B. Tech degree in Computer Science & Engineering from Veer Surendra Sai University of Technology, Odisha, India in 2011 and currently pursuing M. Tech degree in Computer Science & Engineering at National Institute of Technology, Rourkela, Odisha, India.

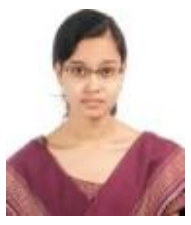

Debashee Tarai has received his B. Tech degree in Computer Science & Engineering from Veer Surendra Sai University of Technology, Odisha, India in 2011 and currently pursuing M. Tech degree in Computer Science & Engineering at Indian Institute of Technology, Bombay, Maharashtra, India.